\documentclass[aip, apl, 10pt, groupedaddress, twocolumn, superscriptaddress]{revtex4-1}
\usepackage[T1]{fontenc}
\usepackage{bm}
\usepackage{graphicx}
\usepackage{amsmath}
\usepackage{amssymb}
\usepackage{sidecap}
\usepackage{dcolumn}
\usepackage{color}
\usepackage{ifpdf}
\usepackage{sidecap}

\begin{document}
\title{\textbf{Graphene membrane as a pressure gauge}}

\author{S. P. Milovanovi\'{c}}\email{slavisa.milovanovic@uantwerpen.be}
\affiliation{Physics Department, University of Antwerp \\
Groenenborgerlaan 171, B-2020 Antwerpen, Belgium}

\author{M. \v{Z}. Tadi\'{c}}\email{milan.tadic@etf.bg.ac.rs}
\affiliation{School of Electrical Engineering, University of Belgrade\\
 P.O. Box 3554, 11120 Belgrade, Serbia}
 
\author{F. M. Peeters}\email{francois.peeters@uantwerpen.be}
\affiliation{Physics Department, University of Antwerp \\
Groenenborgerlaan 171, B-2020 Antwerpen, Belgium}
\begin{abstract}
Straining graphene results in the appearance of a pseudo-magnetic field which alters its local electronic properties. Applying a pressure difference between the two sides of the membrane causes it to bend/bulge resulting in a resistance change. We find that the resistance changes linearly with pressure for bubbles of small radius while the response becomes non-linear for bubbles that stretch almost to the edges of the sample. This is explained as due to the strong interference of propagating electronic modes inside the bubble. Our calculations show that high gauge factors can be obtained in this way which makes graphene a good candidate for pressure sensing. 

\end{abstract}

\pacs{02.60.Cb, 72.80.Vp, 73.23.-b, 75.47.-m}

\date{Antwerp, \today}

\maketitle

Graphene is known as a material with excellent mechanical properties. Experimental studies showed extremely large Young's modulus of $E = 1$ TPa in the case of defect-free graphene membranes with intrinsic strength of $\sigma_{int} = 130$ GPa which is the highest ever measured for real materials \cite{cyoung}. Furthermore, one atom thick graphene membranes proved to be impermeable for all of the standard gases including helium \cite{cmemb01, cmemb02, cmemb03}. This feature of graphene membranes (and its derivatives) was used to construct systems for water filtration and desalination \cite{cgfil01, cgfil02, cgfil03, cgfil04, cgfil05, cgfil06}.

Furthermore, due to the relatively strong van der Waals interaction graphene membranes clamp firmly to a substrate. Koening \textit{et al.} found adhesion energy of 0.45 $\pm$ 0.02 J/m$^2$ for monolayer graphene on SiO$_2$ substrate, which is the most common substrate, and 0.31 $\pm$ 0.03 J/m$^2$ for multilayer graphene membranes \cite{ckoening}. These values are quite large for micromechanical structures and are comparable to solid/liquid adhesion energies. Khestanova \textit{et al.} showed that large adhesion energy allows graphene balloons to sustain extreme pressures. In Ref. \onlinecite{cadhes} they showed that the maximal pressure inside the balloon scales as 1/$h_{max}$, where $h_{max}$ is the maximal out-of-plane deformation, and for balloons with a radius smaller than 10 nm can reach close to 1 GPa.
\begin{figure}[b]
\begin{center}
\includegraphics[width=8.5cm]{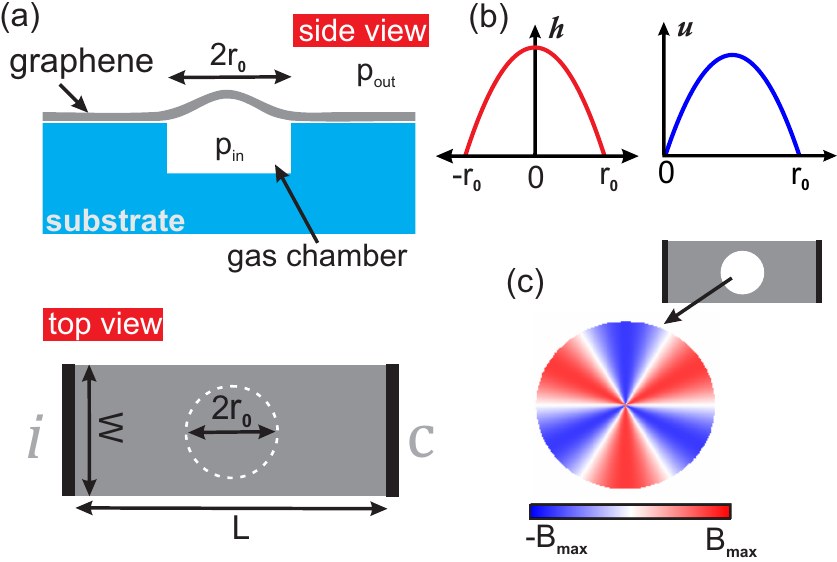}
\caption{(a) Schematics of the system of interest. Graphene stripe is deposited on a substrate with a gas chamber of radius $r_0$ etched in it. The difference in pressure inside and outside the chamber causes bulging/denting of a graphene layer. (b) Profiles of the out-of-plane ($h$) and radial ($u$) displacement obtained using Eq. \eqref{emem1}. (c) The profile of generated PMF obtained using membrane theory. Figures are plotted using: $p = 100$ MPa and $r_0 = 30$ nm.}
\label{f1}
\end{center}
\end{figure}

Another important feature of strained graphene is the occurrence of a pseudo-magnetic field (PMF) for certain strain profiles. The PMF arises as a consequence of the change of the equilibrium positions of the carbon atoms in the crystal structure of graphene. Depending on the profile of the applied strain generated PMF can be quasi-homogeneous \cite{chomo01, chomo02, chomo03, cmass} or inhomogeneous \cite{cinhomo01, cinhomo02, cinhomo03, cinhomo04, cinhomo05} and can well exceed \cite{cextreme} 300 T, much more than "real" magnetic field values that can be realized in laboratory. Important property of the PMF is that it has the opposite direction for electrons in different valleys. Hence, one can use this feature to create a valley filter \cite{cvall01, cvall02, cvall03}, a prerequisite for valleytronics \cite{cvalleytronics01, cvalleytronics02}.
\begin{figure*}[t]
\begin{center}
\includegraphics[width=15.7cm]{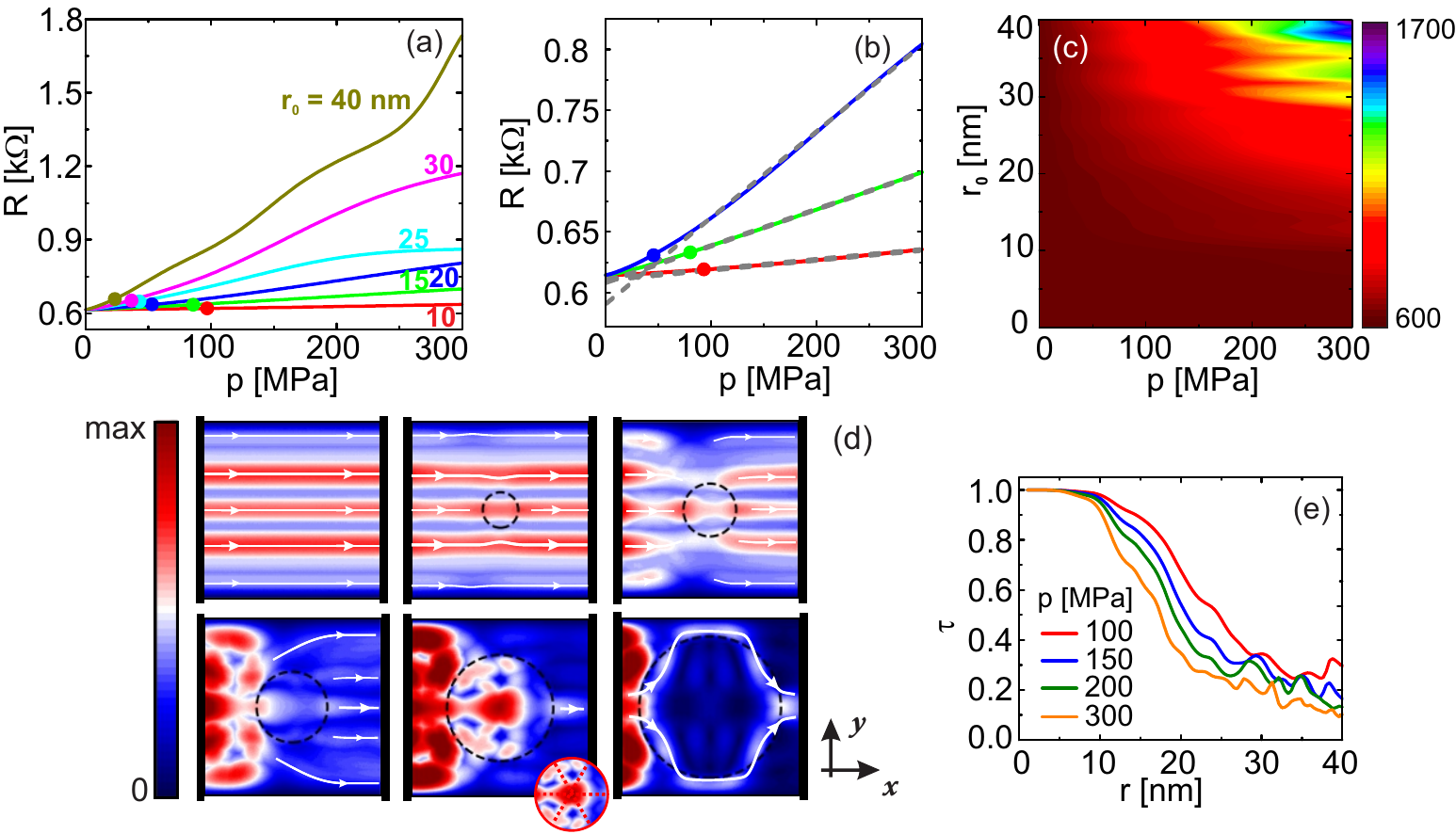}
\caption{(a) Resistance versus applied pressure for different values of bubble radius, $r_0$. Circles show the pressure at which the bubble's height reaches 10 $\%$ of its radius. (b) The resistances for the three smallest bubbles from (a) together with the corresponding linear fits (gray dashed lines). (c) Color plot of $dR/dp$ (in units of $\Omega/$MPa) versus pressure and bubble radius $r_0$. (d) Current intensities for $r_0 = 0, 10, 15, 20, 30, 40$ nm using $p = 200$ MPa and $E_F= 0.05$ eV. (e) Relative change of intramode transmission, $\tau$, versus the radius of the bubble at different applied pressures given in the inset.}
\label{f2}
\end{center}
\end{figure*}

In this paper we propose a pressure sensor using a graphene membrane with electrical read out. Difference in pressures on the two sides of the membrane will cause it to bulge or bend. This generates a pseudo-magnetic field which alters the electronic properties of  graphene. Hence, by observing the change in the conductance/resistance one should be able to estimate the pressure difference between the two sides of the membrane.

Previously, Huang \textit{et al.} investigated the change of resistance in a nanoindentation experiment performed on suspended graphene\cite{cpress_01}. Disappointing results were probably due to the fact that the homogeneous tensile strain used in the experiment does not scatter electrons efficiently. Much better results were obtained in Ref. \onlinecite{cpress_02} where a resistance change from $\sim$ 492 to $\sim$ 522 k$\Omega$ with applied strain up to 1$\%$ was repored. Highly linear resistance response with strain was found in Ref. \onlinecite{cpress_03} using a diamond membrane on a glass substrate as a pressure gauge for application in harsh environments. Experiment showed a sensitivity of $27.3 \pm 0.1$ $\Omega$/bar.

We investigate the change in electrical response of a graphene sample on top of a gas chamber due to the change of pressure inside the chamber. The system of interest is shown in Fig. \ref{f1}(a). A small chamber of radius $r_0$ is etched in a substrate and the pressure inside the chamber, $p_{in}$, is determined by the amount of gas in it. Over the chamber a layer of graphene is deposited. Difference between $p_{in}$ and outside pressure, $p_{out}$, causes the graphene membrane to bulge or dent depending on the value of $p = p_{in} - p_{out}$. Deformation of graphene induces a pseudo-magnetic field (PMF) which affects its electronic properties and will change the resistance of a two-terminal graphene stripe (top view in Fig. \ref{f1}(a)). 
\begin{figure}[b]
\begin{center}
\includegraphics[width=8.5cm]{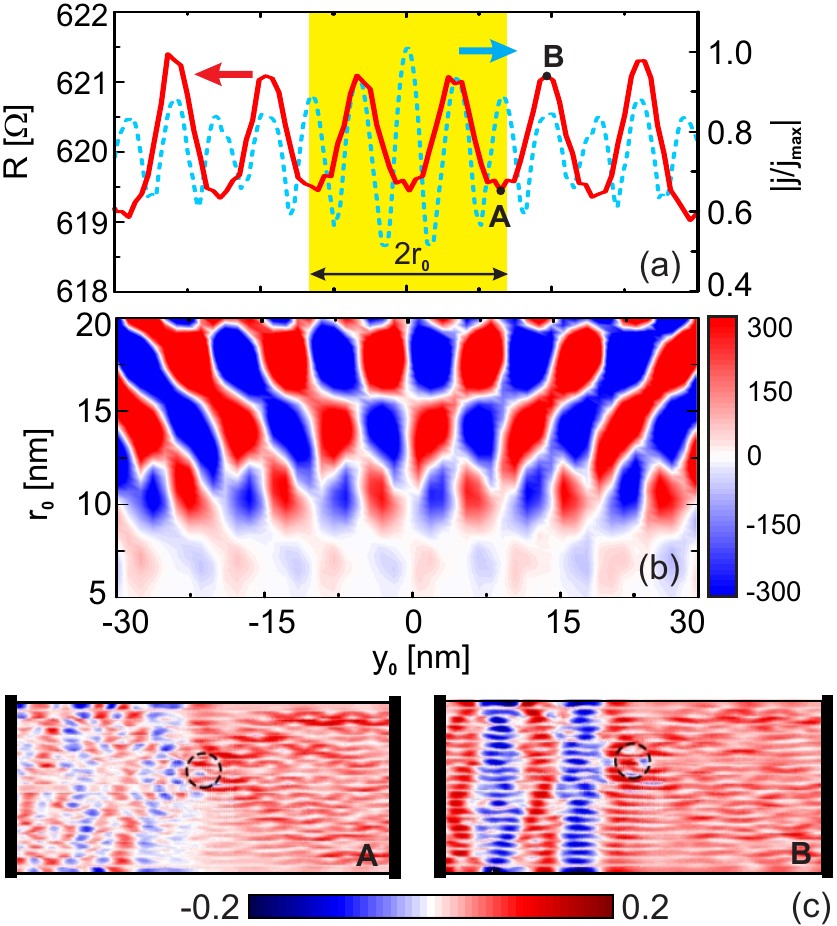}
\caption{(a) Change of resistance with $y_0$ (red curve corresponding to left $y-$axis) using $r_0 = $ 10 nm and normalized current intensity for fixed $x$-coordinate in case when there is no bubble (blue dashed curve corresponding to right $y-$axis). Center of the coordinate system is placed at the center of the structure. (b) $dR/dy_0$ (in units of $\Omega/$nm) versus $y_0$ and $r_0$. (c) Relative change of current intensity, $\tilde{j}$, for two points from (a). In all plots we use $p = 100$ MPa.}
\label{f3}
\end{center}
\end{figure}

Stretching graphene results in changes of the bond length between neighboring atoms in its lattice. This change results in a modification of the hopping energy given by:
\begin{equation}
\label{e1}
t_{ij} = t_0 e^{-\beta(d_{ij}/a_0 - 1)},
\end{equation}
where $t_{ij}$ is the hopping energy between atoms $i$ and $j$, $t_0 = 2.8$ eV is the equilibrium hopping energy, $\beta = 3.37$ is the strained hopping energy modulation factor, $a_0=0.142$ nm is the length of the unstrained $C-C$ bond, and $d_{ij}$ is the length of the strained bond between atoms $i$ and $j$.

The change of hopping energy is equivalent to the generation of a magnetic vector potential, $\mathbf{A} = (A_x, A_y, 0)$, which can be evaluated around the $\mathbf{K}$ point using\cite{cret1},
\begin{equation}
\label{e2}
A_x - \mathtt{i} A_y = -\frac{1}{ev_F}\sum_j \delta t_{ij} e^{\mathtt{i}\mathbf{K}\cdot \mathbf{r_{ij}}},
\end{equation}
where the sum runs over all neighboring atoms of atom $i$, $v_F$ is the Fermi velocity, $\delta t_{ij} = (t_{ij} - t_0)$, and $\mathbf{r_{ij}} = \mathbf{r_i} - \mathbf{r_j}$. In order to properly simulate effects of strain using the tight-binding method we need a correct deformation profile from which we extract $d_{ij}$. Following Refs. \onlinecite{rs01, rs02} we use membrane theory to model the displacement fields. Membrane theory ignores bending stiffness of a strained material and leads to fairly simple analytical expressions for the displacement fields. However, bending stiffness of graphene is rather small\cite{rs03, rs04} which justifies the use of the membrane model. Furthermore, this model shows good agreement with experimental data for large bubbles \cite{rs01, rs11}. For an elastic thin film to be treated as a membrane, the central deflection should be at least several times the film thickness. In the case of graphene, since its thickness is not well defined, the membrane model can be applicable for those bubbles whose height is larger than $10\%$ of its radius \cite{rs01, ckoening, cadhes}. For a bubble of radius $r_0$ the out-of-plane deflection and radial displacement can be written as \cite{rs02},
\begin{equation}
\label{emem1}
h(r) = h_0\left(1 - \frac{r^2}{r_0^2}\right)~~~ \text{and}~~~ u(r) = u_0\frac{r}{r_0}\left(1 - \frac{r}{r_0} \right),
\end{equation} 
where $h_0$ is the maximal out-of-plane displacement and $u_0 = 1.136h_0^2/r_0$. $h_0$ is related to the applied pressure as \cite{rs01}
\begin{equation}
\label{emem3}
h_0 = \left( \frac{p\phi r_0^4}{E_{2D}} \right)^{1/3},
\end{equation} 
where $\phi$ is a function of Poisson's ratio $\nu$ given by $\phi = \frac{75(1-\nu^2)}{8(23 + 18\nu - 3\nu^2)}$ and $E_{2D} \approx 353$ N/m is the 2D Young modulus of graphene \cite{cyoung}. The profile of pseudo-magnetic field obtained using these displacements is shown in Fig. \ref{f1}(b). It has three-fold symmetry with altering regions of positive and negative PMF. Notice that the generated PMF has a jump around $r = r_0$. This is due to a kink (infinite curvature) in the out-of-plane deformation that occurs at $r = r_0$ (see Fig. \ref{f1}(b)).

The change of resistance with applied pressure in case of a $100 \times 200$  nm graphene stripe of Fig. \ref{f1}(a) is investigated. The resistance is calculated according to the Landauer formula given by, $R = \frac{h}{2e^2} \frac{1}{T}$, where $h$ is Planck's constant, $T$ is the transmission probability between two terminals, and the factor 2 is due to the spin degeneracy of the system (valley degeneracy is included in $T$). Transmission probability is calculated using the Pybinding \cite{rpb} and Kwant \cite{rkw} software packages.

The results are shown in Fig. \ref{f2}(a). Calculations were performed for bubbles of different sizes, from small ones ($r_0 = 10$ nm) to the ones that stretch almost to the edges of the stripe ($r_0 = 40$ nm). The circles in this figure indicate pressures at which out-of-plane deflection becomes $10\%$ of the radius of the bubble, i.e. when membrane model becomes valid. Notice that the sensitivity (slope of the curve) increases with the size of the bubble. This is not surprising since a bigger bubble scatters more electrons. However, an important point is that as the size of the bubble increases the resistance curves change behavior from linear to non-linear. Figure shows that for small bubbles ($r_0 \leq 20$ nm) resistance increases linearly with pressure as confirmed by the linear fits given in Fig. \ref{f2}(b). When the size of the bubble increases the response changes to a quadratic function where the non-linear term is 2 orders of magnitude smaller that the linear term. Finally, for bubbles of diameter close to the width of the stripe resistance becomes a non-linear function. This is also confirmed in Fig. \ref{f2}(c) where we plot $dR/dp$ versus the applied pressure and radius $r_0$. Figure shows fairly constant first derivative for $r_0 < 20$ nm implying a linear dependence on pressure. 

The reason for the change from linear to non-linear response can be found in Fig. \ref{f2}(d) where we show current intensity plots for bubbles of different sizes. To see more clearly the effects of the bubble on electrical transport we lowered the Fermi energy in order to decrease the number of propagating modes and increase the spatial separation between them. In case when there is no bubble (shown in the upper left) one can observe five current maxima along the $y$-direction which correspond to five propagating modes that carry current from the injector (left lead) to the collector (right lead). When the small bump is introduced (upper middle plot in Fig. \ref{f2}(d))  the different propagating modes are still visible. Figure shows that only those propagating modes that move around the center of the structure are affected by the bump while the modes that move close to the edges of the sample are almost unaffected by its presence. Increasing the size of the bubble induces mixing of different modes and consequently enhances its influence on current transport. This is seen in the top right part of Fig. \ref{f2}(d) which shows mixed modes around the bubble however, away from the bubble individual modes become apparent again. In the bottom part of Fig. \ref{f2}(d) we show the effect of large bubbles on current flow. In all three cases bubbles are large enough to scatter most of the injected current back to the injector. Interesting is the case of $r_0 = 30$ nm (shown in the bottom middle) where one can notice intense current redistribution inside the bubble. By careful observation we conclude that these features appear around regions where the PMF changes sign (see Fig. \ref{f1}(c)), shown in the inset by red dotted lines. Modes are heavily mixed and we can no longer distinguish them. Reflection and transmission patterns are very complex and hence the transmission probability becomes non-linear. Interestingly, for very large bubbles (bottom right) the current no longer can penetrate through the bubble but flows around it.

The heavy mode mixing is confirmed in Fig. \ref{f2}(e) where we plot the relative change of intramode transmission defined as $\tau = tr(t)/M$, with $tr(t)$ being the trace of the transmission matrix $t$ and $M$ is the number of incoming propagating modes in the injector. Total transmission of the system is given by $T = \sum_{m,n} t_{mn}$ where $t_{mn}$ is the transmission probability between the incoming mode $k_n$ in the injector and the outgoing mode $k_m$ in the collector. Hence, by taking the trace of this matrix we sum the transmission probabilities from each mode $k_n$ from the injector to the outgoing mode $k_n$ in the collector. Naturally, when there is no bubble $\tau$ is unity which means that no scattering occurs. When the bubble is introduced $\tau$ decreases and, as the figure shows, for large bubbles and high pressures $\tau$ drops below $0.1$ indicating heavy mode mixing.

Since modes influence the resistance it would be interesting to see how the resistance changes if we move the center of the bubble, ($x_0, y_0$). From Fig. \ref{f2}(d) it is obvious that only a change of $y$-coordinate of the center, $y_0$, will cause a resistance change since current distribution (in case when there is no bubble) is constant in $x$-direction and, thus, independent of $x_0$. In Fig. \ref{f3}(a) we show the change of resistance with $y_0$ (red curve). Plots are made using $r_0 = 10$ nm and $p = 100$ MPa. One can notice that oscillations of constant period appear. In order to understand this behavior, in the same figure we plot the spatial distribution of normalized current intensity, $|j/j_{max}|$, for the case when there is no bubble (blue dashed curve). Furthermore, with yellow rectangle of width $2r_0$ we highlighted the region occupied by the bubble (if placed at $y_0=0$). Figure shows that extremes occur whenever the center of the bubble is placed on one of the propagating modes. Thus, the output signal is determined by the interference of the modes affected by the bubble. When the modes interfere constructively we have a peak in transmission while reflection increases when the modes interfere destructively. This is seen in Fig. \ref{f3}(c) where we plot relative change of current intensity defined as $\tilde{j}(x, y) = (j^{p=0}(x, y) - j(x, y))/(j^{p=0}(x, y) + j(x, y))$ for two points A and B from Fig. \ref{f3}(a). The interference of different modes is confirmed by the checkerboard pattern shown in Fig. \ref{f3}(b) where we plot $dR/dy_0$ versus $y_0$ and $r_0$. Figure shows that the resistance periodically increases and decreases as a function of both radius of the bubble and the position of the center.
\begin{figure}[htbp]
\begin{center}
\includegraphics[width=8.0cm]{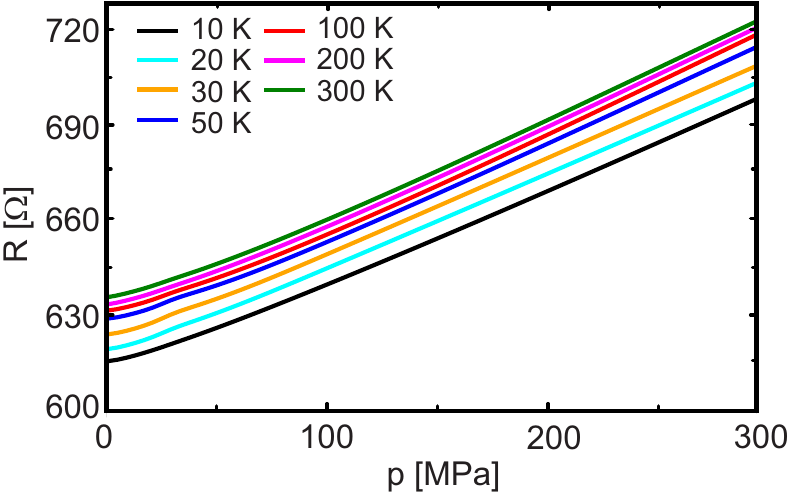}
\caption{Temperature dependence of resistance. Simulations are performed for a bubble with $r_0 = 15$ nm and using a Fermi energy of $E_F = 0.2$ eV.}
\label{f5}
\end{center}
\end{figure}

To compare our results with the experimental results from Refs. \onlinecite{cpress_01} and \onlinecite{cpress_02} we calculated the gauge factor defined as $GF = \Delta R/(R \varepsilon)$, where $\Delta R$ is the change in resistance when maximal straining $\varepsilon$ is achieved. We find $GF = 1.5$ for a bubble with $r_0 = 10$ nm which is comparable to the result of Ref. \onlinecite{cpress_01}. However, for $r_0 = 20$ nm we have $GF \approx 8$ which is closer to the result of Ref. \onlinecite{cpress_02}. For larger bubbles this factor increases further and for $r_0 = 30$ we obtain $GF \approx 18$; however, linearity of the response in this case is lost.

Finally, in Fig. \ref{f5} we show the robustness of our results against temperature.  Simulations are performed for a bubble of radius $r_0 = 15$ nm and $E_F = 0.2$ eV. Increase of temperature leads to an increase of resistance. However, more importantly one can notice that the slope of the curves only slightly change with temperature which indicates that the sensitivity of the device is almost unaffected by it, even up to room temperature.

In conclusion, in this paper we investigated the possibility of using a graphene membrane as a pressure gauge. The results showed that the resistance changes linearly with pressure for bubbles of small radii (as compared to the width of the structure). Increasing the radius of the bubble results in a better sensitivity, however, the linear response is lost in this case. This was explained as due to strong interference of propagating modes inside the bubble. High gauge factors suggest that graphene membranes are good candidates for pressure sensing. We also investigated the influence of temperature on the resistance and found that the resistance increases with temperature and the slope of the $R$-$p$ curve was almost unaffected by it.

This work was supported by the Flemish Science Foundation (FWO-Vl), the Methusalem program, the Erasmus+ programme, and the Serbian Ministry of Education, Science and Technological Development. 

\begin{thebibliography}{99}
%
%
%
%
\bibitem{cyoung} C. Lee, X. Wei, J. W. Kysar, and J. Hone, Science \textbf{321}, 385 (2008).
%
\bibitem{cmemb01} J. S. Bunch, S. S. Verbridge, J. S. Alden, A. M. van der Zande, J. M. Parpia, H. G. Craighead, and P. L. McEuen, Nano Lett. \textbf{8}, 2458 (2008).
%
\bibitem{cmemb02} O. Leenaerts, B. Partoens, and F. M. Peeters, Appl. Phys. Lett. \textbf{93}, 193107 (2008).
%
\bibitem{cmemb03} R. R. Nair, H. A. Wu, P. N. Jayaram, I. V. Grigorieva, and A. K. Geim, Science \textbf{335}, 442 (2012).
%
\bibitem{cgfil01} S. P. Koenig,	L. Wang, J. Pellegrino, and J. S. Bunch, Nat. Nanotech. \textbf{7}, 728 (2012).
%
\bibitem{cgfil02} D. Cohen-Tanugi and J. C. Grossman, Nano Lett. \textbf{12}, 3602 (2012).
%
\bibitem{cgfil03} Y. Han, Z. Xu, and C. Gao, Adv. Funct. Mater. \textbf{23}, 3693 (2013).
%
\bibitem{cgfil04} R. K. Joshi, P. Carbone, F. C. Wang, V. G. Kravets, Y. Su, I. V. Grigorieva, H. A. Wu, A. K. Geim, and R. R. Nair, Science \textbf{343}, 752 (2014).
%
\bibitem{cgfil05} J. Kou, X. Zhou, H. Lu, F. Wu, and J. Fan, Nanoscale \textbf{6}, 1865 (2014).
%
\bibitem{cgfil06} S. P. Surwade, S. N. Smirnov, I. V. Vlassiouk, R. R. Unocic, G. M. Veith, S. Dai, and S. M. Mahurin, Nat. Nanotech. \textbf{10}, 459 (2015).
%
%
\bibitem{ckoening} S. P. Koenig, N. G. Boddeti, M. L. Dunn, and J. S. Bunch, Nat. Nanotech. \textbf{6}, 543 (2011).
%
\bibitem{cadhes} E. Khestanova, F. Guinea, L. Fumagalli, A. K. Geim, and I. V. Grigorieva, Nat. Comm. \textbf{7}, 12587 (2016).
%
\bibitem{cret1} V. M. Pereira and A. H. Castro Neto, Phys. Rev. Lett. \textbf{103}, 046801 (2009).
%
\bibitem{chomo01} F. Guinea, M. I. Katsnelson, and A. K. Geim, Nat. Phys. \textbf{6}, 30 (2009).
%
\bibitem{chomo02} T. Low and F. Guinea, Nano Lett. \textbf{10}, 3551 (2010).
%
\bibitem{chomo03} S. Zhu, J. A. Stroscio, and T. Li, Phys. Rev. Lett. \textbf{115}, 245501 (2015).
%
\bibitem{cmass} M. Ramezani Masir, D. Moldovan, and F. M. Peeters, Solid State Commun. \textbf{175-176}, 76 (2013). 
%
\bibitem{cinhomo01} F. de Juan, A. Cortijo, M. A. H. Vozmediano, and A. Cano, Nat. Phys. \textbf{7}, 810 (2011).
%
\bibitem{cinhomo02} J. Zabel, R. R. Nair, A. Ott, T. Georgiou, A. K. Geim, K. S. Novoselov, and C. Casiraghi, Nano Lett. \textbf{12}, 617 (2012).
%
\bibitem{cinhomo03} T. Mashoff, M. Pratzer, V. Geringer, T. J. Echtermeyer, M. C. Lemme, M. Liebmann, and M. Morgenstern, Nano Lett. \textbf{10}, 461 (2010)
%
\bibitem{cinhomo04} N. N. Klimov, S. Jung, S. Zhu, T. Li, C. A. Wright, S. D. Solares, D. B. Newell, N. B. Zhitenev, and J. A. Stroscio, Science \textbf{336}, 1557 (2012).
%
\bibitem{cinhomo05} D. A. Bahamon, Z. Qi, H. S. Park, V. M. Pereira, and D. K. Campbell, Nanoscale \textbf{7}, 15300 (2015).
%
\bibitem{cextreme} N. Levy, S. A. Burke, K. L. Meaker, M. Panlasigui, A. Zettl, F. Guinea, A. H. Castro Neto, and M. F. Crommie, Science \textbf{329}, 544 (2010).
%
\bibitem{cvall01} M. Settnes, S. R. Power, M. Brandbyge, and A.-P. Jauho, Phys. Rev. Lett. \textbf{117}, 276801 (2016).
%
\bibitem{cvall02} S. P. Milovanovic and F. M. Peeters, Appl. Phys. Lett. \textbf{109}, 203108 (2016).
%
\bibitem{cvall03} R. Carrillo-Bastos, C. Le\'{o}n, D. Faria, A. Latg\'{e}, Eva Y. Andrei, and N. Sandler, Phys. Rev. B \textbf{94}, 125422 (2016).
%
\bibitem{cvalleytronics01} O. Gunawan, B. Habib, E.P. De Poortere, and M. Shayegan, Phys. Rev. B \textbf{74}, 155436 (2006).
%
\bibitem{cvalleytronics02} O. Gunawan, Y. P. Shkolnikov, K. Vakili, T. Gokmen, E. P. De Poortere, and M. Shayegan, Phys. Rev. Lett. \textbf{97}, 186404 (2006). 
%
\bibitem{cpress_01} M. Huang, T. A. Pascal, H. Kim, W. A. Goddard, and J. R. Greer,  Nano Lett. \textbf{11}, 1241 (2011).
%
\bibitem{cpress_02} Y. Lee, S. Bae, H. Jang, S. Jang, S.-E. Zhu, S. H. Sim, Y. I. Song, B. H. Hong, and J.-H. Ahn, Nano Lett. \textbf{10}, 490 (2010).
%
\bibitem{cpress_03} S. D. Janssens, S. Drijkoningen, and K. Haenen, Appl. Phys. Lett. \textbf{104}, 073107 (2014).
%
\bibitem{rs01} K. Yue, W. Gao, R. Huang, and K. M. Liechti, J. Appl. Phys. \textbf{112}, 083512 (2012).
%
\bibitem{rs02} M. Settnes, S. R. Power, J. Lin, D. R. Petersen, and A.-P. Jauho, J. Phys.: Conf. Ser. \textbf{647}, 012022 (2015).
%
\bibitem{rs11} T. Georgiou, L. Britnell, P. Blake, R. V. Gorbachev, A. Gholinia, A. K. Geim, C. Casiraghi, and K. S. Novoselov, Appl. Phys. Lett. \textbf{99}, 093103 (2011).
%
\bibitem{rs03} Q. Lu, M. Arroyo, and R. Huang, J. Phys. D: Appl. Phys. \textbf{42}, 102002 (2009).
%
\bibitem{rs04} P. Koskinen and O. O. Kit, Phys. Rev. B \textbf{82}, 235420 (2010).
%
\bibitem{rpb} D. Moldovan and F. M. Peeters, DOI: 10.5281/zenodo.56818 (2016).
%
\bibitem{rkw} C. W. Groth, M. Wimmer, A. R. Akhmerov, and X. Waintal, New J. Phys. \textbf{16}, 063065 (2014).
%
\end{thebibliography}
\end{document}